\documentclass{svproc}
\usepackage{url}

\usepackage{amsmath,amssymb,amsfonts}
\usepackage{graphicx,lipsum}
\usepackage{subfigure}
\usepackage{adjustbox}
\usepackage{makecell}
\usepackage{tabularx,ragged2e}

\begin{document}
\mainmatter              
\title{Analyzing the Impact of Foursquare and Streetlight Data with Human Demographics on Future Crime Prediction}
\titlerunning{Future Crime Prediction}  
%
\author{Fateha Khanam Bappee\inst{1} \and Lucas May Petry\inst{2} \and
Amilcar Soares \inst{3} \and Stan Matwin \inst{1,4}}
\authorrunning{Fateha Khanam Bappee et al.} 
%
%
\institute{Dalhousie University, Halifax, NS, Canada B3H 4R2,\\
\email{ft487931@dal.ca},
\and
Universidade Federal de Santa Catarina,
Florianópolis, Brazil,
\and
Memorial University of Newfoundland,
St. John’s, NL, Canada
\and
Polish Academy of Sciences, Warsaw, Poland
}

\maketitle              

\begin{abstract}
Finding the factors contributing to criminal activities and their consequences is essential to improve quantitative crime research.
To respond to this concern, we examine an extensive set of features from different perspectives and explanations.
Our study aims to build data-driven models for predicting future crime occurrences.
In this paper, we propose the use of streetlight infrastructure and Foursquare data along with demographic characteristics for improving future crime incident prediction.
We evaluate the classification performance based on various feature combinations as well as with the baseline model.
Our proposed model was tested on each smallest geographic region in Halifax, Canada.
Our findings demonstrate the effectiveness of integrating diverse sources of data to gain satisfactory classification performance.

\end{abstract}
\textbf{NOTICE:} This is a pre-print of a paper with the same name to appear in the 16th International Conference on Data Science (ICDATA’20) proceedings and will be published in Springer Nature - Book Series: Transactions on Computational Science \& Computational Intelligence.  

\section{Introduction}

Crime is one of the well-known social problems that affect the quality of life and slows down the country's economy. 
Nowadays, with the advancement of big data analytics, exploring diverse sources of data has gained increasing interest and attention that offers better crime analysis and prediction for crime researchers. 
Besides, identifying crime patterns and trends is of great importance for police and law enforcement agencies. 
Crime patterns tell us the story about the environment, demography, temporality, and how criminals interact with those factors. 

In this paper, we address the problem of predicting future crime incidents for small geographic areas (i.e., dissemination areas as defined by Statistics Canada) in Halifax, Canada. 
Traditionally, criminology researchers study and analyze the historical crime data by focusing on sociological and psychological theories to obtain crime and criminal behavioral patterns.
However, such strategies may introduce bias from the theory-ladenness of observation \cite{Spatial_Crime_ACM2018}. 
The literature states that in the real world, crime has a mutual relationship with time, space, and population that complicates the researcher's study more \cite{FAR}.
Several works have explored the relationships between criminal activities and socio-economic factors, for instance, educational facilities, ethnicity, income level, unemployment, etc., as well as human behavioral factors \cite{Hojman2004,Graif2009,OUCNuria}. 
Crime rate or crime occurrence prediction has received considerable attention in many studies, including \cite{MTSCorr,BDataCrime,WangPing2010}. 
Several studies tried to predict specific types of crime for a specific region or time by detecting the patterns of that crime \cite{Yu2016}. 
Spatio-temporal pattern plays a vital role in advanced research in crime analysis and prediction \cite{Leong2015}. 
Nowadays, advanced techniques are applied to detect different crime patterns such as spatio-temporal, demographic, meteorological, and human behavioral patterns for crime prediction. 
However, it is challenging to make accurate estimations from diverse data sources due to nonlinear relationships and data dependencies. 

Most of the studies that presented data-driven crime pattern detection and prediction approaches have focused on mega-cities like Chicago, New York, Greater London, etc. 
However, the physical characteristics, human impact characteristics, and their interactions are totally different for different regions and cities. 
Therefore, applying those models for predicting crime in a smaller city is challenging and may lead to different outcomes.
Our study aims to build data-driven models for future crime incidents prediction for smaller cities. The main hypothesis is that the relative scarcity of data, compared to mega-cities, can be compensated by using non-traditional datasets that can be derived from social media and the Internet-of-Things (IoT) infrastructure of a modern city. 
We extract five different categories of features from six different data sources. 
We propose to explore traditional demographics data with commuting features (e.g., commuting mode and time), IoT-like streetlight poles position data, as well as human mobility data with dynamic features from location-based social networks. 
To the best of our knowledge, employing demographics data with human mobility features for future crime prediction is the first attempt for a small city such as Halifax, Canada. 
For model building, we use ensemble learning methods such as Random Forest and Gradient Boosting. 
We conduct a performance comparison for all five categories of features. 
We also compare the prediction results generated from ensemble learning methods with a baseline method called DNN-based feature level data fusion \cite{DLMMD17}.  

In summary, the contributions of this paper are:
(i) we propose the use of streetlight infrastructure data with demographic characteristics for improving future crime prediction. 
Its effectiveness is demonstrated in our experimental evaluation results;
(ii) we propose data-driven models to predict future crime occurrences in smaller cities. 
This implies that fewer data points are applicable for training the models;
and 
(iii) we experimentally show the effect of each feature group proposed in previous works and this paper on crime prediction, evaluating the classification performance of different feature combinations.

The rest of the paper is organized as follows. Section \ref{sec:relatedWork} reviews the related work.
Section \ref{sec:Feat} provides the details of feature engineering approaches to improve the prediction accuracy in Halifax.
After, in Section \ref{sec:Experiments}, the data source, data preparation activities, and experimental results are presented. 
Finally, Section \ref{sec:conclusions} presents some concluding remarks with future research directions.

\section{Related Work}
\label{sec:relatedWork}

The relationship between crime and various factors has been studied in many scientific and criminology works. 
Nowadays, researchers can use spatial information from the real world using Geographic Information Systems (GIS).
Likewise, demographic information is easily accessible from different statistical sources. 
The use of historical facts and temporal dynamics between neighborhoods and crimes have also been broadly noted in criminology. 
Researchers have emphasized the feasible computation solutions for the urban crime after analyzing the factors related to different categories of crimes and their consequences. 
We have categorized the existing work of crime prediction from four aspects: temporal and historical, geographic, demographic and streetlight, and human behavioral aspects.

Temporal patterns of crime are learned from sequential crime data by analyzing the structure (various intervals) of temporal resources. 
Crime rates can be examined for hours of the day, different days of the week, months, seasons, years, and others. 
Many researchers have studied how to identify temporal patterns among criminal incidents \cite{DLMMD17,CNNST17}. 
Several works also focus on historical information to predict future crime incidents \cite{Yu2011}.
In \cite{DLCF17,DLCFT17}, the authors presented a periodic temporal pattern with hourly crime intensity and holiday information for crime forecasting. 
A study \cite{Yu2016} shows that drunk driving incidents and other criminal incidents occur during Saturday nights, bar game nights close to the bar, and sports season close to the stadium. This implies that the temporal influence of crime may change over geographic regions.

Existing works also examined the geographic influence for future crime prediction or crime rate estimation \cite{BDataCrime}.  
Wang et al. \cite{BDataCrime} presented a crime rate inference problem for Chicago community areas by utilizing Point-of-Interest (POI) data as well as geographical influence features. 
Geospatial Discriminative Patterns (GDPatterns) was introduced in \cite{WangDawei2013} to capture the spatial properties of crime. 
Furthermore, spatial autocorrelation is considered in \cite{Yu2011} where the average number of neighbors is calculated for each grid. Besides, the authors in \cite{Almanie2015,AlBoni2017} found spatial patterns (hotspots) for crime prediction using the Apriori algorithm and Localized Kernel Density Estimation (LKDE), respectively.
Recently, another study \cite{Bappee18} focused on the creation of spatial features to predict crime using geocoding and crime hotspots techniques.
As the distributions of crime vary in time and space, several studies have identified spatio-temporal patterns for crime prediction \cite{Fitterer2015,MTSCorr}. 
In \cite{Fitterer2015}, the authors investigated a spatio-temporal dynamic for Break and Entries (BNEs) crime incidents.
However, considering the geographic influence may add a little help for crime prediction as the neighboring community shares similar demographics.

Traditional demographic features have been extensively used in many research for crime prediction \cite{OUCNuria,DLMMD17,Spatial_Crime_ACM2018}. 
In \cite{FAR}, the author applied population density, mean people per household, people in the urban area, people under the poverty level, and people in dense housing with some other features to detect community crime patterns. 
A study discovered the association of construction permits, foreclosures, etc. with crime tendency \cite{Mu2011}. 
Researchers also explored residential stability, number of vacant houses, number of people who are married or separated, and education \cite{BDataCrime,Spatial_Crime_ACM2018}.
However, using only traditional demographic feature is insufficient to understand the implicit characteristics of crime and criminals. 
Few works reported the impact of streetlight distributions on the criminal behavioral pattern and crime prediction. The researchers in 2018 \cite{stlight2018} have found an inverse relationship between streetlight density and crime rates based on the census block groups in Detroit. 
In our study, we also consider extracting streetlight features, but for crime incidents prediction. 
However, due to human mobility, a region's demographic characteristics may change for a short or long period of time.

Human behavioral pattern aims to obtain understanding from human behavior, mobility, and networks. 
In \cite{OUCNuria}, the authors investigated the predictive power of aggregated and anonymized human behavioral data derived from a multimodal combination of mobile network activity and demographic information. 
Specifically, footfall or the estimated number of people within each cell is derived from the mobile network by aggregating every hour the total number of unique phone calls in each cell tower and mapping the cell tower coverage areas to the Smartsteps cells. 
Similar works have been done by Andrey et al. \cite{Bogomolov2015} and Traunmueller et al. \cite{socinfo} for crime hotspots classification and to find a correlation between crime and metrics derived from population diversity. 
In \cite{BDataCrime}, the authors profile the crime rate by applying taxi flow data to understand the reflection of city dynamics.
The authors considered taxi flows as `hyperlinks' in the city to connect the locations.
Each taxi trip recorded pick-up/dropoff time and location, operation time, and the total amount paid. 
The taxi flow features indicate how neighboring areas contribute much crime in the target area through social interaction.
A data-driven approach is presented in \cite{Spatial_Crime_ACM2018} for crime rate prediction that also considers road network, transportation nodes, and human mobility.
Recently, crime event prediction for Brisbane and New York are studied in \cite{Kadar2016,Rumi2018} using dynamic features extracted from foursquare data. The authors measure the region's popularity by determining the total number of observed check-ins in that region for a specific time interval. Also, the number of unique users that checked in to a specific venue and the number of tips users have ever written about that venue are counted to measure the popularity, heterogeneity, and quality of the region.   

In our study, we proposed a data-driven approach for a smaller city, Halifax, by investigating an extensive set of features from all different aspects. 
We mainly focus on human behavioral aspects, streetlight features, and the traditional demographic features for future crime prediction.

\section{Feature Extraction}
\label{sec:Feat}

Aiming at predicting future crime incidents, we extract features for each dissemination area (DA). 
According to Statistics Canada, a DA is the smallest standard geographic area in their data, which consists of one or more adjacent dissemination blocks~\cite{disseminationA}.
This section is organized as follows.
Section \ref{subsec:temp_hist_feats} details the temporal and historical features used in this work. 
The demographics and streetlight features are explained in Section \ref{subsec:demog_light_feats}, while Section \ref{subsec:poi_feats} shows the POI features used in this work. 
Finally, Section \ref{subsec:dyn_feats} shows some human mobility dynamic features extracted from social networks.


\subsection{Temporal and Historical Features}
\label{subsec:temp_hist_feats}

According to criminology research, crime may change over a long period of time (e.g., season) as well as in a short period of time (e.g., day or week) \cite{Ratcliffe_2004}. 
Thus, the temporal features we extracted are month, day of the week, time interval in a day, and season.
We arrange crime records in 8 three-hour time intervals and 4 seasons (winter, fall, summer, and spring) for each DA.
On the other hand, some research analyzed the relation of future crime incidents with the past crime history \cite{Yu2011}.
Therefore, we calculate crime frequency and crime density for each region based on historical crime data.
As the area and population sizes are different for different regions, we normalize the crime frequency using the area and population size to obtain the crime density ($D_{cr}$).

\begin{equation}
    D_{cr}(r) = \frac{CR(r)}{P(r)},
\end{equation}
\begin{equation}
    D_{cr}(r) = \frac{CR(r)}{A(r)},
\end{equation}
where $CR(r)$ addresses the total number of crimes in DA $r$, $P(r)$ is the total number of population in region $r$, and $A(r)$ is the area of that region. We also compute the crime distribution based on each season.

\subsection{Demographic and Streetlight Features}
\label{subsec:demog_light_feats}

Researchers have widely used demographic and socioeconomic features for crime rate estimation \cite{BDataCrime} and crime occurrence prediction \cite{OUCNuria}. 
The main demographic features we consider for our study are population density, dwelling characteristics, income, mobility, the journey to work, aboriginals and visible minorities, age, and sex.
The journey to work features measure two main things: (i) the time people leave for work and (ii) the primary mode of commute for residents aged more than 15 years. 
We consider 6 different measures for the time people leave for work, such as between 5 a.m. and 5:59 a.m, 6 a.m. and 6:59 a.m., 7 a.m. and 7:59 a.m., 8 a.m. and 8:59 a.m., 9 a.m. and 11:59 a.m., and 12 p.m. and 4:59 a.m.
For the mode of commute, public transit, walk, bicycle, and other methods are considered.  
Mobility indicates the geographic movement of a population over a period of time, for instance, it shows the information if a person moved to the current place of residence or is living at the same place as 1 year or 5 years ago. 

Besides demographic features, we observe the effect and graveness of streetlight distribution on future crime incidents prediction motivated by \cite{stlight2018}.
Given a dataset of streetlight locations, for each DA, we propose the use of 3 streetlight features: (i) the total number of streetlights, (ii) the streetlight density, and (iii) the average minimum distance between crime data points and streetlight poles. The streetlight density of region $r$ is computed as follows:

\begin{equation}
    D_{st}(r) = \frac{St(r)}{A(r)},
\end{equation}
where $St(r)$ denotes the total number of streetlights in DA $r$. 
To calculate the average minimum distance from crime location to streetlight poles, we use the Haversine distance metric with scikit-learn~\cite{scikit-learn}. 

\begin{figure}
    \centering
    \subfigure[Crime density.]
    {
        \includegraphics[width=.30\textwidth]{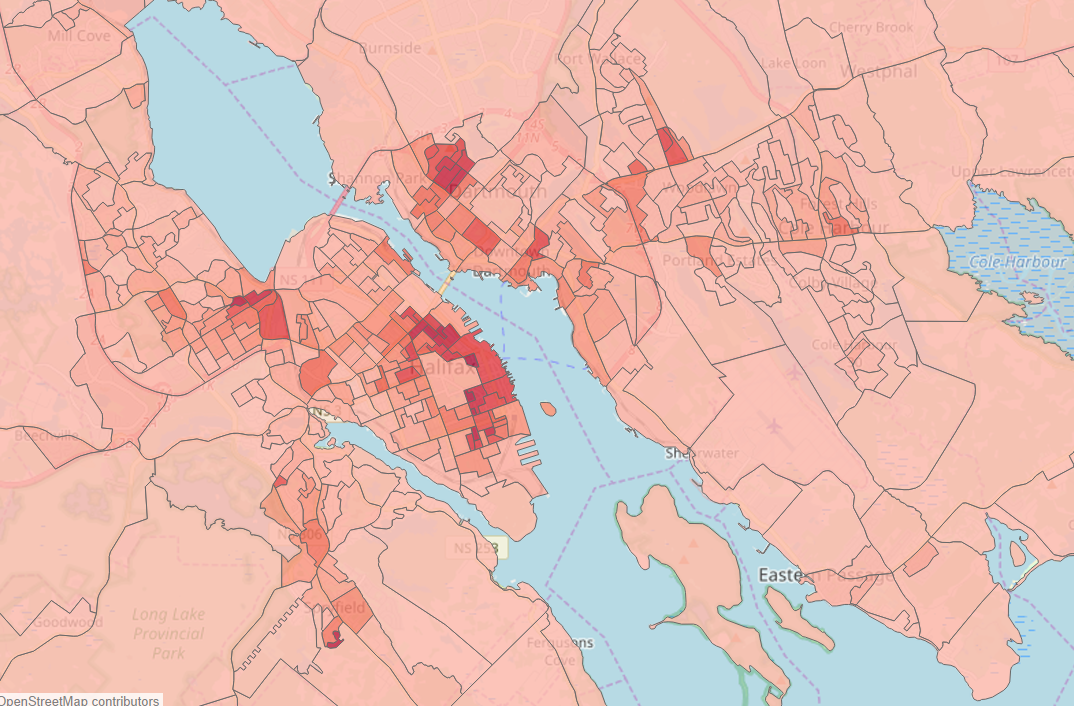}
        \label{fig:first_sub}
    }\hspace{.5em}
    \subfigure[Population density.]
    {
        \includegraphics[width=.30\textwidth]{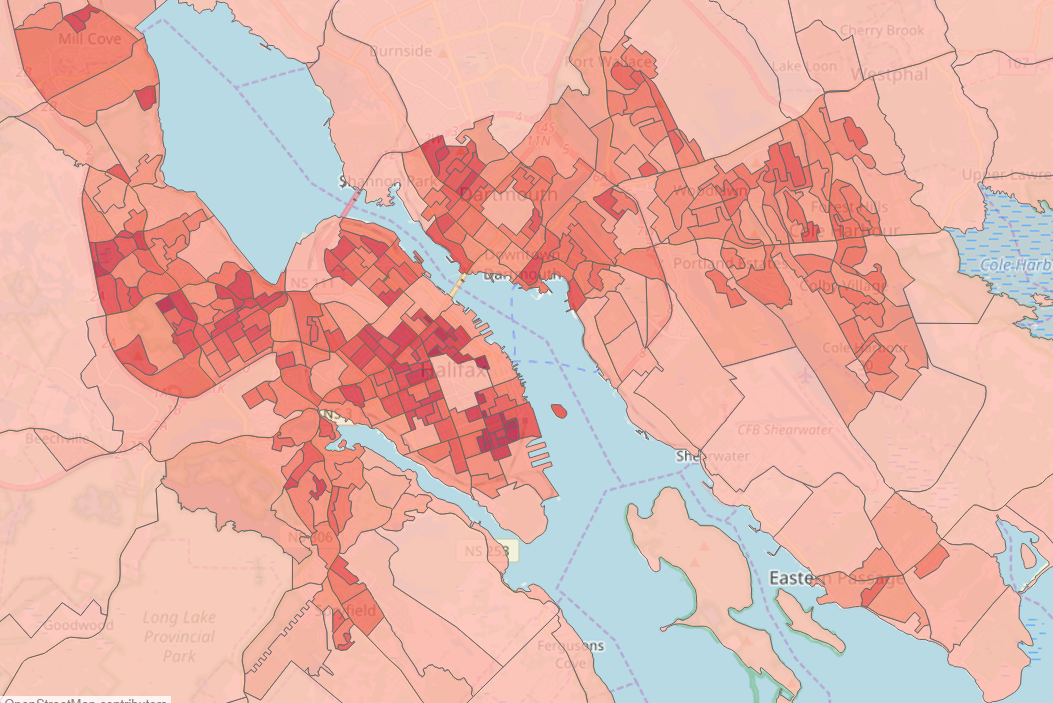}
        \label{fig:second_sub}
    }\hspace{.5em}
    \subfigure[Streetlight density.]
    {
        \includegraphics[width=.30\textwidth]{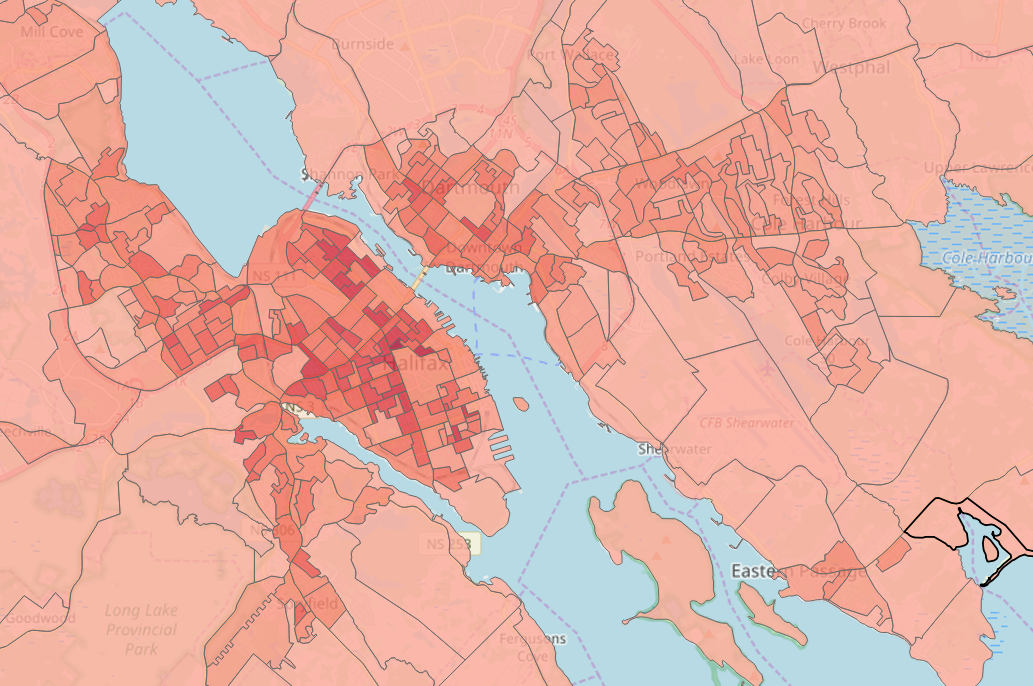}
        \label{fig:third_sub}
    }
    \caption{Crime, population density and streetlight density by most observable DAs in Halifax.}
    \label{fig:sample_subfigures}
\end{figure}

Figure \ref{fig:sample_subfigures} visualizes the crime (year 2013), population, and streetlight densities by most observable DAs in Halifax. 
Dark red color indicates high density, and light red indicates low density. 
The bin sizes for population and streetlight densities are the same; on the other hand, we choose smaller bin sizes to get a clear picture of crime density. As shown in Figure \ref{fig:sample_subfigures}, most of the criminal incidents happen in downtown Halifax.   

\subsection{POI Features}
\label{subsec:poi_feats}

\begin{figure}
    \centering
    \subfigure[Poi count distribution.]
    {
        \includegraphics[width=.4\textwidth]{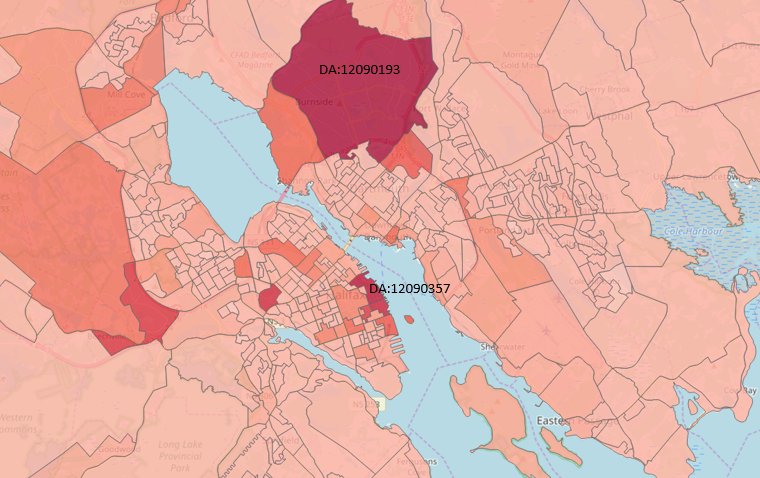}
        \label{fig:poi}
    }\hspace{.5em}
    \subfigure[Check-in count distribution.]
    {
        \includegraphics[width=.4\textwidth]{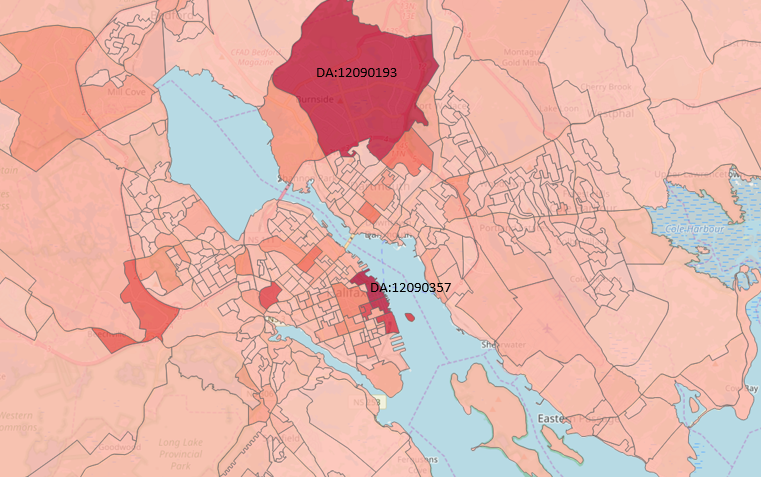}
        \label{fig:ch-in}
    }
    \caption{The total POI and check-in count distributions by most observable DAs in Halifax.}
    \label{fig:poi_ch}
\end{figure}

In this work, we propose the use of POI features that can be obtained from location-based social networks (e.g., Foursquare). 
Our extracted POI features include (i) the total number of POIs, (ii) the POI frequency, and the density for different POI categories. Foursquare identifies 10 major POI categories, such as food, arts and entertainment, college and university, nightlife spots, outdoors, and recreation, professional and other places, residence, shop, and service event, and travel \& transport. 
The density of each POI category is defined as follows:

\begin{equation}
    D_c(r) = \frac{N_c(r)}{N(r)},
\end{equation}
\begin{equation}
    {D_1}_c(r) = \frac{N_c(r)}{A(r)},
\end{equation}

where, $N_c(r)$ is the total number of POIs of category $c$ in a DA $r$, $N(r)$ is the total number of POIs in region $r$, and $A(r)$ is the area of that region. 
Figure \ref{fig:poi_ch} shows the POI and check-in count distributions of most observable dissemination areas (DAs) in Halifax. 

\subsection{Human Mobility Dynamic Features}
\label{subsec:dyn_feats}

Our study also explores dynamic human mobility data from location-based social networks to find if there is any relation with crime context.
Social networks often have their users' location data, including their visits to different POIs in a city.
We extract 10 features for each DA based on the total number of user check-ins and check-in frequency for each POI category.
Moreover, the check-in count for each DA at a time interval, the check-in density, region popularity, and visitor count are also computed. 
For DA $r$ at time interval $t$, the check-in density is defined as follows:

\begin{equation}
    D_{ck}(r,t) = \frac{Ck(r,t)}{Ck(r)},
\end{equation}
\begin{equation}
    D_{ck}(r,t) = \frac{Ck(r,t)}{A(r)},
\end{equation}

where, $Ck(r,t)$ is the number of check-ins in DA r at time interval t, and $Ck(r)$ is the total number of check-ins in that region.
Visitor count refers to the number of unique users that visited a DA at time interval $t$ (i.e., region popularity). 

\begin{equation}
    R_{rp}(r,t) = \frac{Ck(r,t)}{Ck(t)},
\end{equation}

where, $Ck(t)$ is the total number of check-in at time interval t for all regions. 

We extract a total of 153 features for each dissemination area. Among them, we select 65 features that are more relevant to the crime prediction problem. The details of the total features chosen for each category appear in Table \ref{fset}.

\begin{table*}[!h]
\centering
\caption{Details of the selected features}
\resizebox{\columnwidth}{!}{%
\begin{tabular}{l|c|c|c}
\hline
\textbf{Feature category} & \textbf{Extracted features} & \textbf{Selected features} & \textbf{Selected feature names} \\
\hline
Temporal and historical & 12 & 8 & \makecell{Month, weekday, time interval, season, crime frequency, \\ crime density based on population, crime density based on area, \\ crime density for season} \\
\hline
Demographic & 101 & 32 & \makecell{Population, population density, dwelling characteristics (11) \\ mobility movers, mobility non movers, mobility migrants,\\ mobility non migrants, aboriginals and visible minorities, \\primary mode of commute for residents (5), journey to work: \\the time people leave for work (5), low income (3), age and sex} \\
\hline
Streetlight & 3 & 2 & \makecell{streetlight frequency, streetlight density} \\
\hline
Foursquare POI & 21 & 19 & \makecell{Total POI, food count, residence count, nightlife count, \\arts \& entertainment count, college \& University count, \\outdoors \& recreation count, professional \& other places count, \\shop \& service count, travel \& transport count, and \\the densities of all POI categories (9)} \\
\hline
Foursquare dynamic & 16 & 4 & \makecell{Total check-in for each time interval, check-in density, visitor count, \\region popularity} \\
\hline
\end{tabular}\label{fset}}
\end{table*}

\section{Experiments}
\label{sec:Experiments}

We conducted experiments to evaluate the effectiveness of the different groups of features that can be aggregated to crime data for the task of crime prediction.
In the following sections, we describe the datasets used (Section~\ref{subsec:datasets}), the experimental setup (Section~\ref{subsec:exp_setup}), and the achieved results (Section~\ref{subsec:exp_results} and \ref{subsec:exp_results_baseline}).

\subsection{Datasets}
\label{subsec:datasets}

We use crime data provided by the Halifax Regional Police (HRP) department, which includes records for all Dissemination Areas (DAs) in the Halifax Regional Municipality (HRM) in Nova Scotia, Canada.
Our dataset was extracted from the Uniform Crime Reporting (UCR) survey, which was designed to measure the incidence of crime and its characteristics in Canadian society.
For our experiments, we explore all crime occurrences from 2012 to 2014.
The crime attributes extracted from the dataset include the geographic location, incident start time, month, weekday, and UCR descriptions (incident type).
We have a total of 201,086 crime observations (excluding invalid and null information), where 69,340 data points happened in 2012, 65,785 in 2013, and 65,961 in 2014.
We map all crime records to one of the 599 DAs collected for Halifax from statistics Canada 2016 census, based on their geographic location.
We group and index crime occurrences based on the DA where they happened, the year, month, day of the week, and the time interval of the day (we partition a day into 8 three-hour time intervals).

In addition to the raw crime data, we collected demographic data for each DA from the Canadian census analyser~\cite{demof}.
We also extracted POI and dynamic features for Halifax from a dataset of Foursquare check-ins around the world, collected between April 2012 and January 2014 \cite{yang2016participatory}.
Lastly, streetlight information was obtained from the Streetlight Vision (SLV) API of HRM, which contains the location of 42,653 streetlight poles after removing null values and invalid data.
We then computed the streetlight features proposed in Section~\ref{sec:Feat} and mapped them to each DA.

Given that there are only records of crime occurrences in the dataset, we augment it to include 'no crime' records.
Thus, if there was no crime for a specific time interval, we labeled that observation as 'no crime'.
The final size of the dataset, including crime and no crime records, is 1,207,584 (3*12*7*8*599).
 
As the occurrence of crime events is not frequent, most of the data (around 87\%) are labeled with 'no crime'.
To address this issue, we apply the under-sampling technique for 'no crime' records to obtain a more balanced dataset \cite{JMLR:v18:16-365}. 

We use the random under-sampling technique, which randomly selects a subset of observations from the major class (no crime) of the dataset.
Applying random under-sampling might lead to a biased dataset; also, the deleted data points could have a beneficial or adverse impact to fit the model.
However, this under-sampling approach is compatible for our study as we are employing it for artificially creating `no crime' records only. The number of records for `crime' occurrences is sufficient despite having class imbalance.   
Table \ref{dset} shows the details of the dataset. 


\begin{table}[htb]
\centering
\caption{Details of the datasets}
\begin{tabular}{l|l|c}
\hline
\textbf{Dataset} & \textbf{Source} & \textbf{Total data} \\
\hline
Historical crime data & Halifax Regional Police & 201,086 \\
\hline
Dissemination area data & Statistics Canada & 599 \\
\hline
Demographic data & Canadian Census Analyser & 599 \\
\hline
Streetlight data & Halifax Regional Municipality & 42,653 \\
\hline
Foursquare POI data & Foursquare & 2,301 \\
\hline
Foursquare checkin data & Foursquare & 12,171 \\
\hline
\end{tabular}\label{dset}
\end{table}




\subsection{Experimental setup}
\label{subsec:exp_setup}

We run experiments with well-known ensemble learning classifiers, Random Forest (RF)~\cite{breiman2001random} and Gradient Boosting (GB)~\cite{friedman2001greedy}, with scikit-learn~\cite{scikit-learn} in Python.

We used randomized grid-search in preliminary experiments for the hyper-parameter optimization of each classifier evaluated.
Besides evaluating the effect of each group of features, we compare our results to a DNN-based feature level data fusion baseline method~\cite{DLMMD17}.
Since the environmental context feature group used in \cite{DLMMD17} is unavailable for Halifax, we implement the DNN without those features.
We use the same parameter settings reported in the corresponding paper for the baseline model, except for the activations of the DNN, which were replaced by sigmoid functions as they resulted in a better performance.
We train the DNN for 300 epochs and select the best test scores.
For evaluating the effectiveness of each feature group, we analyze the accuracy and F-score of the classifiers. 
At the same time, for the comparison with the baseline method, we also report precision, recall, and Area Under the ROC Curve (AUC) scores.

We run a 10-fold time-constrained cross validation, similar to what was proposed in \cite{DBLP:journals/corr/abs-1905-11744}.
This is more appropriate since we guarantee that the records in the training set happened before the ones used for testing, and so we are effectively using data from the past to predict the future.
We consider a sliding time window of 2 years, where the first 12 months are taken for training the models, and the subsequent 12 months are used for testing.
Thus, the models are still capable of capturing seasonality patterns as the training split always contains a full year of data. 
As our dataset includes three consecutive years of crime records from 2012 to 2014, for the first fold, we take all records from January 2012 to December 2012 for training, and the test split goes from January 2013 to December 2013.
Next, for the second fold, we slide the window one month forward so that the training set spans from February 2012 to January 2013, and the test spans from February 2013 to January 2014.
We repeat this process for 10 different folds.

\subsection{Results for our proposed features}
\label{subsec:exp_results}

In Table \ref{tab:acc}, we show the classification results with various feature combinations. 
We tested the addition of four different groups of features to the Raw dataset (temporal + historic crime) (R): Demographic (D), Streetlight (S), Foursquare dynamic (F), and Foursquare POI (P) features.
We compare 12 different models by adding all feature categories one by one with the raw features.
Our first model is implemented based on the raw features only, named as model MR.
We built models MD, MS, MF, and model MP by adding demographic, streetlight, foursquare check-in, and foursquare poi data, respectively, with the raw data. 
Similarly, by combining two consecutive feature groups with the raw data, we built the models MDS, MDF, MDP, MSF, MSP, and model MFP.
Finally, model MA is implemented based on all of the feature combinations. 
Both RF and GB classifiers share a similar trend for all models based on classification accuracy and F-score. 
As GB performs better than RF for all combinations, in our discussions, we only consider the GB method. 
Model MR, trained only with raw features, is resulting in low accuracy of 59.85\% and 64.65\% F-score.
Such behavior is expected since criminal behavior is affected by many different variables other than simple spatial and temporal factors \cite{DBLP:journals/corr/abs-1804-08159}.

By analyzing the addition of each group of features individually (top part of Table~\ref{tab:acc}), the inclusion of demographic features (model MD) exhibits the best results, for which GB shows an improvement of almost 10\% in accuracy (69.94\%) and about 5\% in F-score (69.45\%) compared to only raw features.
Similarly, streetlight features in model MS show an approximate 9\% and 4\% improvement for accuracy and F-score, respectively.
Demographic variables reveal most of the characteristics of different regions, including social and economic factors, which are commonly correlated with criminality.
Likewise, the installment of streetlight poles that reflects streetlight density feature also considers the same demographic profile for each corresponding area.
Interestingly, Foursquare POI and dynamic features achieve less accuracy individually as compared to demographic and streetlight features.
One of the reasons for this may be that there is missing information for places and check-in data for some dissemination areas.

\begin{table*}[!h]
\centering
\caption{Results for average accuracy and F-score}
\label{tab:acc}
\begin{tabular}{c|c|c|c|c|c|c|c|c|c|c}
\cline{1-11}
                                      \null   & \null   &               \multicolumn{5}{c|}{\textbf{Features}} &            \multicolumn{2}{c|}{\textbf{Random Forest}} & \multicolumn{2}{c}{\textbf{Gradient boosting}} \\ \hline
\multicolumn{1}{c|}{\textbf{No.}} & \multicolumn{1}{c|}{\textbf{Model}}      & \textbf{R}   & \textbf{D}   & \textbf{S}   & \textbf{F}   & \textbf{P}    & \textbf{Accuracy (\%)}   & \textbf{F-score (\%)}      & \textbf{Accuracy (\%)} & \textbf{F-score (\%)}   \\ \hline
\multicolumn{1}{c|}{1} & \multicolumn{1}{c|}{MR} & \checkmark   & \null  & \null  &\null  &\null  & 59.60            & 63.93       & 59.85       & 64.65    \\ \hline
\multicolumn{1}{c|}{2} & \multicolumn{1}{c|}{MD} & \checkmark   & \checkmark  & \null  &\null  &\null  & 69.07               & 68.30      & \textbf{69.94}       & \textbf{69.45}    \\ \hline
\multicolumn{1}{c|}{3} & \multicolumn{1}{c|}{MS} & \checkmark   &\null & \checkmark   &\null  &\null  & 68.51               & 67.46      & 68.52       & 68.25    \\ \hline
\multicolumn{1}{c|}{4} & \multicolumn{1}{c|}{MF} & \checkmark   &\null  &\null  & \checkmark  &\null  & 64.08               & 61.25     & 64.70       & 61.16    \\ \hline
\multicolumn{1}{c|}{5} & \multicolumn{1}{c|}{MP} & \checkmark   &\null  &\null  &\null       & \checkmark   & 66.75               & 64.19      & 67.61       & 64.19    \\ \hline \hline
\multicolumn{1}{c|}{6} & \multicolumn{1}{c|}{MDS} & \checkmark   &\checkmark  &\checkmark  &\null       & \null   & 69.08               & 68.32      & 69.97       & \textbf{69.51}    \\ \hline
\multicolumn{1}{c|}{7} & \multicolumn{1}{c|}{MDF} & \checkmark   &\checkmark  &\null  &\checkmark       & \null   & 69.15               & 68.29      & 69.95       & 69.33    \\ \hline
\multicolumn{1}{c|}{8} & \multicolumn{1}{c|}{MDP} & \checkmark   &\checkmark  &\null  &\null       & \checkmark   & 68.98               & 68.20      & \textbf{70.00}       & 69.42    \\ \hline
\multicolumn{1}{c|}{9} & \multicolumn{1}{c|}{MSF} & \checkmark   &\null  &\checkmark  &\checkmark       & \null   & 68.06               & 66.48      & 69.04       & 67.50    \\ \hline
\multicolumn{1}{c|}{10} & \multicolumn{1}{c|}{MSP} & \checkmark   &\null  &\checkmark  &\null       & \checkmark   & 68.66               & 66.89      & 69.50       & 68.21    \\ \hline 
\multicolumn{1}{c|}{11} & \multicolumn{1}{c|}{MFP} & \checkmark   &\null  &\null  &\checkmark       & \checkmark   & 66.84               & 64.02      & 67.60       & 64.07    \\ \hline \hline
\multicolumn{1}{c|}{12} & \multicolumn{1}{c|}{MA} & \checkmark   &\checkmark  &\checkmark  &\checkmark       & \checkmark   & 69.09               & 68.16      & \textbf{69.96}       & \textbf {69.31}   \\ \hline
\end{tabular}
\end{table*}

Models 6 to 11 show the evaluation results for three feature categories combination. The accuracy and F-scores are better and almost consistent for all models except model MFP. 
The reason behind this is that all of them contain either demographic or streetlight features except MFP.  
In model MA, we combine all five categories of features. 
It gives us similar results as Model 6 (MDS), where we added both the demographic and streetlight categories. 
As Foursquare features do not lead to performance loss while combining others, in our study, we used all feature categories for building a model. 


\subsection{Comparison with a baseline}
\label{subsec:exp_results_baseline}

Table \ref{bmodel2} reports the accuracy, precision, recall, and AUC score for one of the best performing ensemble-based models, Model MA with Gradient Boosting (GB-MA) and the baseline DNN model.
Our proposed model performs significantly better than the baseline model based on precision, recall, and AUC scores.
Though DNN can handle non-linear relationships and data dependencies among different sources, it is challenging for the model to perform accurately for smaller domains or domains that suffer from data scarcity. 
This is the most likely reason for the baseline model to degrade performance.
On the other hand, our model performed, on average, about 2\% worse considering accuracy compared to the baseline model.
This is due to the existence of a label imbalance in some of the testing folds.

\begin{table}[htbp]
\centering
\caption{Performance evaluation}
\begin{tabular}{l|c|c|c|c}
\hline
\textbf{Model} & \textbf{Accuracy (\%)} & \textbf{Precision (\%)} & \textbf{Recall (\%)} &\textbf{AUC}\\
\hline
DNN (baseline) \cite{DLMMD17} & \textbf{71.82} & 49.52 & 49.74 & 50.00 \\
\hline
GB-MA & 69.96 & \textbf{70.13} & \textbf{68.53} & \textbf{69.95} \\
\hline
\end{tabular}
\label{bmodel2}
\end{table}

\section{Conclusions and Future Work}
\label{sec:conclusions}

In this paper, we study a fundamental problem of crime incidents prediction for the future time interval. 
We have presented a data-driven approach to see how prediction accuracy can be improved by integrating multiple sources of data. 
Specifically, we focus on exploring population-centric features with streetlight and Foursquare-based features for each dissemination area in Halifax. 
Our problem also considers the temporal dimension of the crime profile in depth. 
We compare all 5 categories of feature combinations differently and unitedly. 
The results show that demographic and streetlight features have strong correlations with crime. 
Both of them show significant performance improvement for crime prediction individually and jointly. 
Though Foursquare data does not outperform demographic or streetlight data, it presents a satisfactory performance for crime prediction. 
Additionally, we compare our best ensemble model (i.e., Model MA with Gradient Boosting in Table \ref{tab:acc}) with the DNN-based baseline model.
Our results show that GB outperforms the DNN baseline for the same groups of features. 
Therefore, applying ensemble-based method leads to better performance in predicting future crime for smaller cities, such as Halifax.


In the future, we plan to extend this work in multiple directions. We want to integrate real-time streetlight data (e.g., light temperature, lux level, outages determined by power supply failures, etc.) with the current dataset. 
Moreover, identifying specific types of crime that might happen in the near future is our immediate concern. 
As it is very challenging to get accurate results for future crime prediction when sufficient data is unavailable, performing domain adaptation and some form of transfer learning using available data from a big city, would be advantageous. 
Furthermore, the subject of data discrimination is another crucial concern for the study that focuses on real-world datasets. 
Investigating discrimination in socially-sensitive decision records is state-of-the-art research to avoid biased classification learning. 

\bibliographystyle{./bibtex/spmpsci}
\bibliography{./main}

\end{document}